\documentclass[journal ]{new-aiaa}
\usepackage[utf8]{inputenc}
\usepackage{textcomp}
\usepackage{hyperref}
\hypersetup{
    colorlinks=true,
    linkcolor=blue,
    filecolor=magenta,      
    urlcolor=cyan,
    pdftitle={Overleaf Example},
    pdfpagemode=FullScreen,
    }
\usepackage{graphicx}
\usepackage{amsmath}
\usepackage[version=4]{mhchem}
\usepackage{siunitx}
\usepackage{longtable,tabularx}
\title{Sedimentation of a surfactant-laden drop in a liquid with particles}

\author{Xiaoxu Zhong \footnote{Ph.D. student. Email: zhong150@purdue.edu}}
\affil{School of Mechanical Engineering, Purdue University, West Lafayette, IN 47906, United States}

\begin{document}

\maketitle

\begin{abstract}
This project aims to study the sedimentation of a surfactant-laden drop in a liquid with particles. A 2D simulation is performed with MATLAB. The interface is captured by the front-tracking method. The local viscosity depends on the local particle concentration, which follows a power law. The surfactants not only decrease the surface tension but also induce a surface tension gradient.  If the surface tension decreases, the settling velocity will decrease due to a larger deformation of the drop (i.e., the drop becomes more flat).
Additionally, increasing the Peclet number (ratio of convection to diffusion on the interface) for surfactants will reduce the settling velocity due to a larger surface tension gradient. 
\end{abstract}

\section{Introduction}
\lettrine{T}{he} settling of a drop in a liquid is a classic problem in fluid mechanics. The terminal velocity of a rigid sphere settling through a liquid with viscosity $\eta$ and density $\rho$ is known as $U_{rigid}=2(\rho'-\rho)g R^2/(9\eta)$, in which $\rho'$, $R$, and $g$ are the drop density, drop radius, and gravitational acceleration, respectively.  For a clean spherical drop with viscosity $\eta'$ and radius $R$, its terminal velocity is $U = U_{rigid}(\lambda+1)/(\lambda + 2/3)$, in which $\lambda = \eta'/\eta$ is the viscosity ratio \citep{Manikantan2020S}.  

This project aims to study the settling of a deformable drop in a liquid with surfactants and particles. The particles distribute in both the drop and the outside liquid with negligible particle-particle interaction.  The surfactants are insoluble and they only distribute at the interface.  As the drop settles, the surfactants will migrate to the top of the drop due to advection, thus, induce a gradient of surface tension (known as Marangoni effect).  Also, the viscosity "felt" by the drop changes due to the change of the particle concentration.  For the sake of simplicity, this project neglects the surface rheology of the drop and performs 2D simulation.

\section{Model}
\begin{figure}[hbt!] 
\centering
\includegraphics[width=0.6\textwidth]{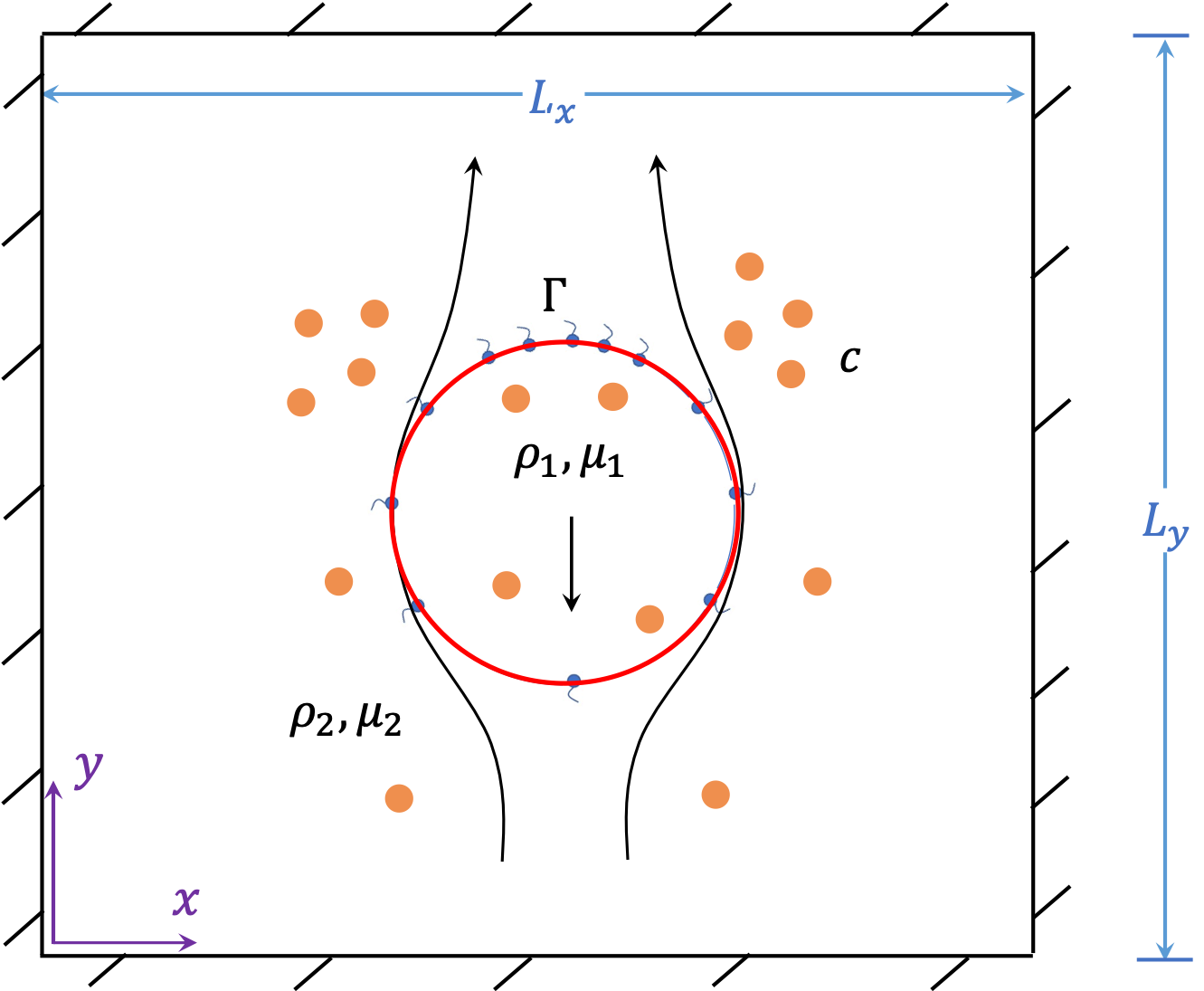} 
\caption{Schematic of a drop settles down in a liquid.} \label{figure:schematic}
\end{figure}

The schematic is shown in Figure \ref{figure:schematic}. The density and viscosity of the drop (outside liquid) are $\rho_1$ ($\rho_2$) and $\mu_1$ ($\mu_2$), respectively. The concentration of surfactants at the interface is $\Gamma$.  The particle concentration is $c$. Following assumptions are taken in this project:
\begin{itemize}
  \item Particles are soluble in both the drop and the outside liquid. The diffusion coefficient of particles in the drop $D_1$ is the same as that in the outside liquid $D_2$, i.e., $D_1 = D_2 = D_c$.
  \item $\mu_1$ and $\mu_2$ are the function of $c$, which follows a power law, i.e., $\mu_1 = \mu_1^{(0)}(1-c/c_{max})^{-2}$,  $\mu_2 = \mu_2^{(0)}(1-c/c_{max})^{-2}$. $\mu_1^{0} = \mu_2^{(0)}=\mu^{(0)}$, $c_{max}$ is the maximum density of particles.  
  \item Surfactants only distribute at the interface. The surface rheology is neglected. 
\end{itemize}

The flow is incompressible which reads
\begin{equation}
    \nabla \cdot \Bar{U} = 0,
\end{equation}
where $\Bar{U} = (u,v)$ is the velocity field. The density field $\rho = \rho_1 \chi + \rho_2 (1-\chi)$, where $\chi = 1$ inside the drop, $\chi = 0$ outside the drop.  $\dot{\rho}=0$ yields
\begin{equation}
    \frac{\partial{\chi}}{\partial t} + \Bar{U} \cdot \nabla \chi = 0.
\end{equation}
The conservation of momentum reads
\begin{equation}
    \rho\frac{\partial{\Bar{U}}}{\partial t} + \nabla \cdot (\rho\Bar{U} \otimes \Bar{U}) = -\nabla p + \mu\nabla^2\Bar{U} + \rho\Bar{g} + \gamma \kappa \Bar{n}\delta_s(\Bar{n}),
\end{equation}
where $\gamma$ and $\kappa$ are the surface tension and curvature, respectively.  The conservation of surfactant concentration reads \citep{Stone1990A}
\begin{equation}
    \frac{\partial{\Gamma}}{\partial t} + \nabla_s \cdot (\Gamma \Bar{U}) - D_{\Gamma}\nabla_s^2 \Gamma = 0,
\end{equation}
where $D_{\Gamma}$ is the diffusion coefficient of the surfactants along the interface, $\nabla_s$ is the surface gradient. The conservation of particle concentration reads
\begin{equation}
    \frac{\partial{c}}{\partial t} + \nabla \cdot (\Bar{U} c) = D_c\nabla^2 c.
\end{equation}
The surface tension varies with the surfactant concentration, which is described by the Langmuir-relation \citep{Manikantan2020S}
\begin{equation}
    \gamma = \gamma_0 + \Gamma_{\infty} R_g T \ln{(1-\Gamma / \Gamma_{\infty})},
\end{equation}
where $\Gamma_{\infty}$ is the maximum surfactant concentration, $R_g$ is the gas constant, $T$ is the temperature.  The boundary conditions read
\begin{eqnarray}
&\;& u = v = 0,\qquad \partial{p}/\partial{x} = \partial{c}/\partial{x} =  0, \qquad \mbox{on $x = 0, L_x$}, \\
&\;& u = v = 0, \qquad \partial{p}/\partial{y} = \partial{c}/\partial{y} =  0, \qquad \mbox{on $y = 0, L_y$}.
\end{eqnarray}

The governing equations, boundary conditions, and the surface equation of state are rendered dimensionless using following scaling:
\begin{eqnarray}
    &\;&(L_x, L_y) = L_x(1, \tilde{L}_y), \quad (u, v) = V(\tilde{u}, \tilde{v}), \quad (\rho_1, \rho_2) = \rho_2(\tilde{\rho}_1, 1), \quad t = \tilde{t}L_x/V, \\
    &\;& c = c_{max}\tilde{c}, \quad p = \rho V^2 \tilde{p}, \quad g = \tilde{g}V^2/L_x, \quad  \Gamma = \Gamma_\infty \tilde{\Gamma}, \quad \gamma = \gamma_0 \tilde{\gamma}.
\end{eqnarray}
Here, $V=\rho_2 g L_x^2 / \mu^{(0)}$ is the characteristic velocity.  The dimensionless governing equations and the surface equation of state read (discard tildes henceforth)
\begin{eqnarray}
&\;&\nabla \cdot \Bar{U} = 0, \\
&\;&\frac{\partial{\chi}}{\partial t} + \Bar{U} \cdot \nabla \chi = 0, \\
&\;&\rho\frac{\partial{\Bar{U}}}{\partial t} + \nabla \cdot (\rho\Bar{U} \otimes \Bar{U}) = -\nabla p + \frac{1}{ \mathrm{Re}}(1-c)^{-2}\nabla^2\Bar{U} +\rho \Bar{g} + \Omega \gamma \kappa \Bar{n}\delta_s(\Bar{n}), \\
&\;&\frac{\partial{\Gamma}}{\partial t} + \nabla_s \cdot (\Gamma \Bar{U}) - \frac{1}{\mathrm{Pe}_\Gamma}\nabla_s^2 \Gamma = 0, \label{dimensionless:Gamma}\\
&\;&\frac{\partial{c}}{\partial t} + \nabla \cdot (\Bar{U} c) = \frac{1}{\mathrm{Pe}_c}\nabla^2 c,\\
&\;&\gamma = 1 + \omega \ln{(1-\Gamma)},
\end{eqnarray}
where $\omega = \Gamma_{\infty} R_g T / \gamma_0 $ represents the surfactant activity, $\Omega = \gamma_0 / (\rho_2 L_x V^2)$, $\mathrm{Re}=\rho_2 V L_x / \mu^{(0)}$, $\mathrm{Pe}_c = VL_x / D_c$, $\mathrm{Pe}_{\Gamma} = VL_x / D_{\Gamma}$.

\section{Discretization}
Staggered grids are used here. Pressure $p_{i,j}$, particle concentration $c_{i,j}$, density $\rho_{i,j}$, and characteristic function $\chi_{i,j}$ are placed at the center of cells, $u_{i,j}$ and $v_{i,j}$ lie on the vertical and horizontal boundaries, respectively. The surface of the drop is captured by the front-tracking method.  The front points are denoted by $x_{f,1}$, $x_{f,2}$, $...$, $x_{f,N(t)}$.  The surfactant concentration $\Gamma_{k}$ is placed at the center of the front points $x_{f,k}$ and $x_{f,k+1}$.  Let $\Delta t$ denote the temporal step size, $\Delta x$ and $\Delta y$ represent the spatial step sizes along $x-$ and $y-$ directions, respectively.    The equation (\ref{dimensionless:Gamma}) can be rewritten as \citep{Muradoglu2008A}
\begin{equation}
    \frac{d(\Gamma A)}{dt} = \frac{A}{\mathrm{Pe}_{\Gamma}}\nabla_s^2\Gamma = \frac{A}{\mathrm{Pe}_{\Gamma}} \frac{d^2 \Gamma}{d S^2},
\end{equation}
where $A$ is the area (arc length in 2D) of an element of the interface, $S$ is the arc length along the interface.  The projection method is applied with following procedures:

\begin{figure}[hbt!] 
\centering
\includegraphics[width=0.8\textwidth]{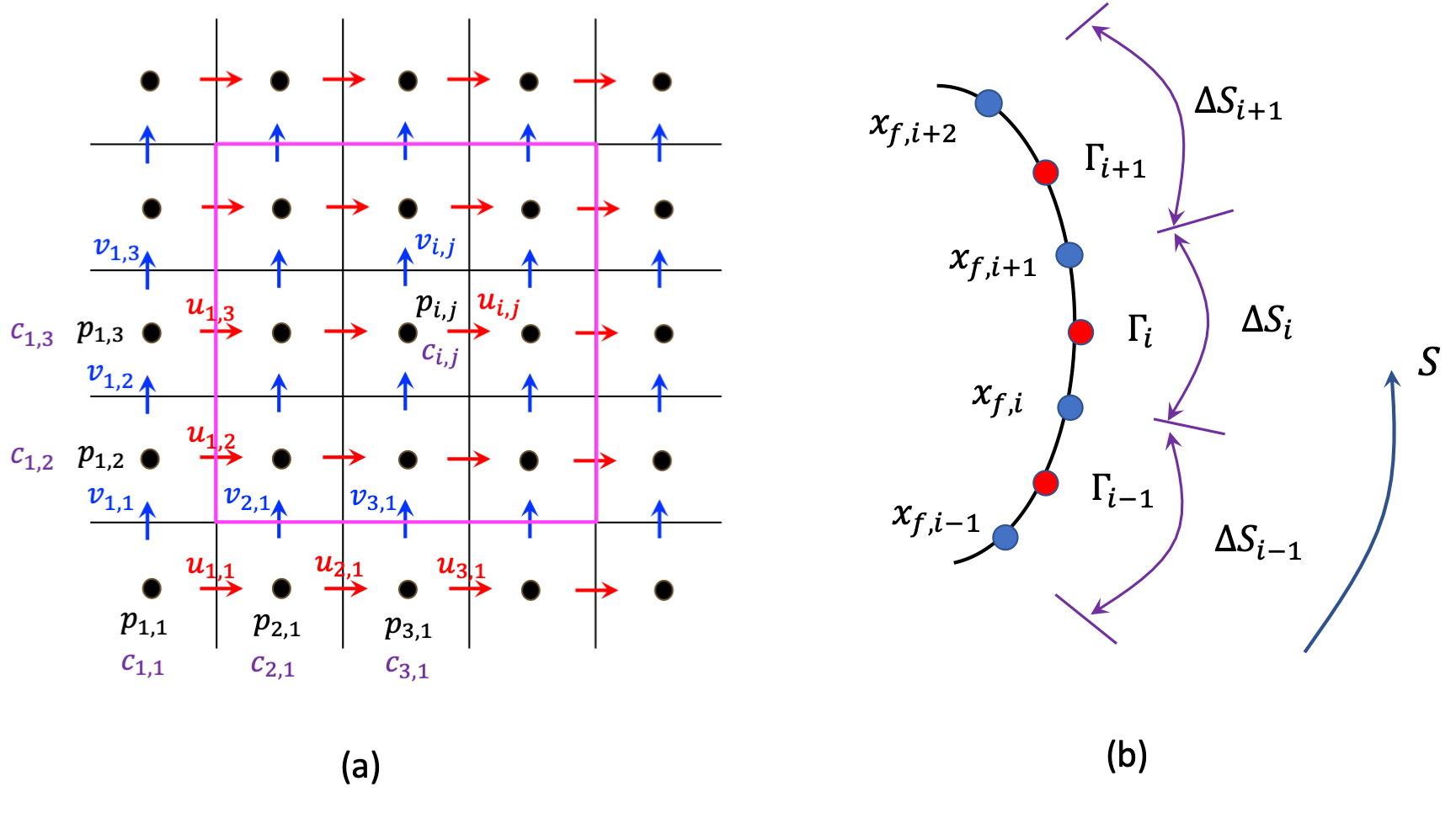} 
\caption{(a) Staggered Grids. (b) Grids on the interface.} \label{figure:staggeredGrid}
\end{figure}

\begin{enumerate}
    \item Compute $c_{i,j}^{(n+1)}$ according to
    \begin{eqnarray}
    \frac{c_{i,j}^{(n+1)} - c_{i,j}^{(n)}}{\Delta t} + \frac{1}{\Delta x}\left(u_{i,j}^{(n)} \frac{c_{i,j}^{(n)} + c_{i+1,j}^{(n)}}{2} - u_{i-1,j}^{(n)} \frac{c_{i-1,j}^{(n)} + c_{i,j}^{(n)}}{2} \right) + \frac{1}{\Delta y}\left(v_{i,j}^{(n)} \frac{c_{i,j}^{(n)} + c_{i,j+1}^{(n)}}{2} - v_{i,j-1}^{(n)} \frac{c_{i,j}^{(n)} + c_{i,j-1}^{(n)}}{2} \right)  \nonumber \\ 
     = \frac{1}{\mathrm{Pe}_c}\left(\frac{c_{i+1,j}^{(n)} - 2 c_{i,j}^{(n)} + c_{i-1,j}^{(n)}}{\Delta x^2} + \frac{c_{i,j+1}^{(n)} - 2 c_{i,j}^{(n)} + c_{i,j-1}^{(n)}}{\Delta y^2} \right);
    \end{eqnarray}
    
    \item Obtain the velocity of front points $U_{f,k}^{(n)}$ by interpolation. Update the position of front points, $x_{f,k}^{(n+1)} = x_{f,k}^{(n)} + U_{f,k}^{(n)} \Delta t$, the characteristic function $\chi_{i,j}^{(n+1)}$, and the arc length between front points, $\Delta S_{k}^{(n+1)} = || x_{f,k+1}^{(n+1)} - x_{f,k}^{(n+1)} ||$;
    
    \item Compute $\Gamma_{k}^{(n+1)}$ according to
    \begin{equation}
        \frac{\Gamma_k^{(n+1)} \Delta S_k^{(n+1)} - \Gamma_k^{(n)} \Delta S_k^{(n)} }{\Delta t} = \frac{2}{\mathrm{Pe}_{\Gamma}}\left(\frac{\Gamma_{k+1}^{(n)}-\Gamma_k^{(n)}}{\Delta S_k^{(n)} + \Delta S_{k+1}^{(n)}} - \frac{\Gamma_{k}^{(n)}-\Gamma_{k-1}^{(n)}}{\Delta S_k^{(n)} + \Delta S_{k-1}^{(n)}} \right);
    \end{equation}
    
    \item Update density $\rho^{(n+1)} = 1 + (\rho_1-1) \chi_{i,j}^{(n+1)}$;
    
    \item Calculate the surface tension $\gamma_f^{(n+1)} = \Omega \delta_{s}(\Bar{n}_s) d(\gamma \Bar{t})/{d S}$ and distribute the force to the grid points, $\gamma_{f,i,j}^{(n+1)}$;
    
    \item Compute $u_{i,j}^*$ according to
    \begin{eqnarray}
    &\;&\frac{\rho_{i,j}^{(n+1)} + \rho_{i+1,j}^{(n+1)}}{2}\cdot \frac{u_{i,j}^{*} - u_{i,j}^{(n)}}{\Delta t}  \nonumber\\
    &=&- \frac{1}{\Delta x}\left(\rho_{i+1,j}^{(n+1)}\frac{u_{i,j}^{(n)} + u_{i+1,j}^{(n)}}{2} \cdot \frac{u_{i,j}^{(n)} + u_{i+1,j}^{(n)}}{2} - \rho_{i,j}^{(n+1)}\frac{u_{i-1,j}^{(n)} + u_{i,j}^{(n)}}{2}\cdot \frac{u_{i-1,j}^{(n)} + u_{i,j}^{(n)}}{2} \right) \nonumber \\
    &\;&- \frac{1}{\Delta y}\bigg{(}\frac{\rho_{i,j}^{(n+1)} + \rho_{i+1,j}^{(n+1)} + \rho_{i,j+1}^{(n+1)} + \rho_{i+1,j+1}^{(n+1)} }{4}\cdot \frac{u_{i,j}^{(n)} + u_{i,j+1}^{(n)}}{2} \cdot\frac{v_{i,j}^{(n)} + v_{i+1,j}^{(n)}}{2} \nonumber\\
    &\;&- \frac{\rho_{i,j}^{(n+1)} + \rho_{i+1,j}^{(n+1)} + \rho_{i,j-1}^{(n+1)} + \rho_{i+1,j-1}^{(n+1)} }{4} \cdot \frac{u_{i,j}^{(n)} + u_{i,j-1}^{(n)}}{2}\cdot \frac{v_{i,j-1}^{(n)} + v_{i+1,j-1}^{(n)}}{2} \bigg{)} \nonumber \\
    &\;& + \frac{1}{\mathrm{Re}} \left(1-\frac{c_{i,j}^{(n+1)} + c_{i+1,j}^{(n+1)}}{2} \right)^{-2}\left(\frac{u_{i+1,j}^{(n)} - 2 u_{i,j}^{(n)} + u_{i-1,j}^{(n)}}{\Delta x^2} + \frac{u_{i,j+1}^{(n)} - 2 u_{i,j}^{(n)} + u_{i,j-1}^{(n)}}{\Delta y^2} \right)+ \gamma_{f,i,j,x}^{(n+1)} ;
    \end{eqnarray}
    
    \item Compute $v_{i,j}^*$ according to
    \begin{eqnarray}
    &\;&\frac{\rho_{i,j}^{(n+1)} + \rho_{i,j+1}^{(n+1)}}{2}\cdot \frac{v_{i,j}^{*} - v_{i,j}^{(n)}}{\Delta t}  \nonumber\\
    &=&- \frac{1}{\Delta x}\bigg{(}\frac{\rho_{i,j}^{(n+1)} + \rho_{i+1,j}^{(n+1)} + \rho_{i,j+1}^{(n+1)} + \rho_{i+1,j+1}^{(n+1)} }{4}\cdot \frac{u_{i,j}^{(n)} + u_{i,j+1}^{(n)}}{2} \cdot\frac{v_{i,j}^{(n)} + v_{i+1,j}^{(n)}}{2} \nonumber\\
    &\;&- \frac{\rho_{i,j}^{(n+1)} + \rho_{i-1,j}^{(n+1)} + \rho_{i,j+1}^{(n+1)} + \rho_{i-1,j+1}^{(n+1)} }{4} \cdot \frac{u_{i-1,j}^{(n)} + u_{i-1,j+1}^{(n)}}{2}\cdot \frac{v_{i,j}^{(n)} + v_{i-1,j}^{(n)}}{2} \bigg{)} \nonumber \\
    &\;&- \frac{1}{\Delta y}\left(\rho_{i,j+1}^{(n+1)}\frac{v_{i,j}^{(n)} + v_{i,j+1}^{(n)}}{2} \cdot \frac{v_{i,j}^{(n)} + v_{i,j+1}^{(n)}}{2} - \rho_{i,j}^{(n+1)}\frac{v_{i,j-1}^{(n)} + v_{i,j}^{(n)}}{2}\cdot \frac{v_{i,j-1}^{(n)} + v_{i,j}^{(n)}}{2} \right) + \frac{\rho_{i,j}^{(n+1)} + \rho_{i,j+1}^{(n+1)}}{2}g_y \nonumber \\
    &\;& + \frac{1}{\mathrm{Re}} \left(1-\frac{c_{i,j}^{(n+1)} + c_{i,j+1}^{(n+1)}}{2} \right)^{-2}\left(\frac{v_{i+1,j}^{(n)} - 2 v_{i,j}^{(n)} + v_{i-1,j}^{(n)}}{\Delta x^2} + \frac{v_{i,j+1}^{(n)} - 2 v_{i,j}^{(n)} + v_{i,j-1}^{(n)}}{\Delta y^2} \right) + \gamma_{f,i,j,y}^{(n+1)} ;
    \end{eqnarray}
    
    \item Update $p_{i,j}$ according to $\nabla \cdot U^* / \Delta t = \nabla \cdot (\nabla p / \rho)$, whose discretization is
    \begin{eqnarray}
    &\;&\frac{1}{\Delta t}\left(\frac{u_{i,j}^* - u_{i-1,j}^*}{\Delta x} + \frac{v_{i,j}^* - v_{i,j-1}^*}{\Delta y} \right) \nonumber \\
    &\;&= \frac{1}{\Delta x}\left(\frac{2}{\rho_{i,j}^{(n+1)} + \rho_{i+1,j}^{(n+1)}} \cdot \frac{p_{i+1,j}^{(n+1)}- p_{i,j}^{(n+1)}}{\Delta x} - \frac{2}{\rho_{i-1,j}^{(n+1)} + \rho_{i,j}^{(n+1)}}\cdot \frac{p_{i,j}^{(n+1)}- p_{i-1,j}^{(n+1)}}{\Delta x} \right) \nonumber\\
    &\;& +\frac{1}{\Delta y}\left(\frac{2}{\rho_{i,j}^{(n+1)} + \rho_{i,j+1}^{(n+1)}} \cdot\frac{p_{i,j+1}^{(n+1)}- p_{i,j}^{(n+1)}}{\Delta y} - \frac{2}{\rho_{i,j}^{(n+1)} + \rho_{i,j-1}^{(n+1)}}\cdot \frac{p_{i,j}^{(n+1)}- p_{i,j-1}^{(n+1)}}{\Delta y} \right);
    \end{eqnarray}
    
    \item Compute $u_{i,j}^{(n+1)}$ and $v_{i,j}^{(n+1)}$ according to
    \begin{eqnarray}
    u_{i,j}^{(n+1)} = u_{i,j}^{*} - \frac{2\Delta t}{\rho_{i+1,j}^{(n+1)} + \rho_{i,j}^{(n+1)}} \cdot  \frac{p_{i+1,j}^{(n+1)} - p_{i,j}^{(n+1)}}{\Delta x}, \\
    v_{i,j}^{(n+1)} = v_{i,j}^{*} - \frac{2\Delta t}{\rho_{i,j+1}^{(n+1)} + \rho_{i,j}^{(n+1)}}\cdot \frac{p_{i,j+1}^{(n+1)} - p_{i,j}^{(n+1)}}{\Delta y};
    \end{eqnarray}
    
    \item Add or delete points in the front if the distance between adjacent points is too large or too small. The surfactant concentration $\Gamma$ will also be adjusted when adding or deleting the front points to guarantee the conservation of mass of surfactants at interface.  
\end{enumerate}

The mass center ($x_c$, $y_c$) of the drop is obtained by
\begin{equation}
    x_c = \frac{\Sigma \chi_{i,j} x_{i,j}}{\Sigma \chi_{i,j}}, \qquad y_c = \frac{\Sigma \chi_{i,j} y_{i,j}}{\Sigma \chi_{i,j}}
\end{equation}

\section{Results}
$L_y = L_x$, $g_y = -100$, density ratio $\rho_1 / \rho_2 = 2$, $\mathrm{Re} = 100$, the Peclet number for particles $\mathrm{Pe}_c = 100$, and the surface activity $\omega = 1$ are used in following analysis. Initially, the drop is spherical with radius $r_c = 0.15$, $x_c^{(0)} = 0.5$, $y_c^{(0)} = 0.7$; The surfactants are uniformly distributed at the interface. $80\times 80$ uniform meshes with $\Delta t = 10^{-5}$ are used. 

\begin{figure}[hbt!] 
\centering
\includegraphics[width=0.8\textwidth]{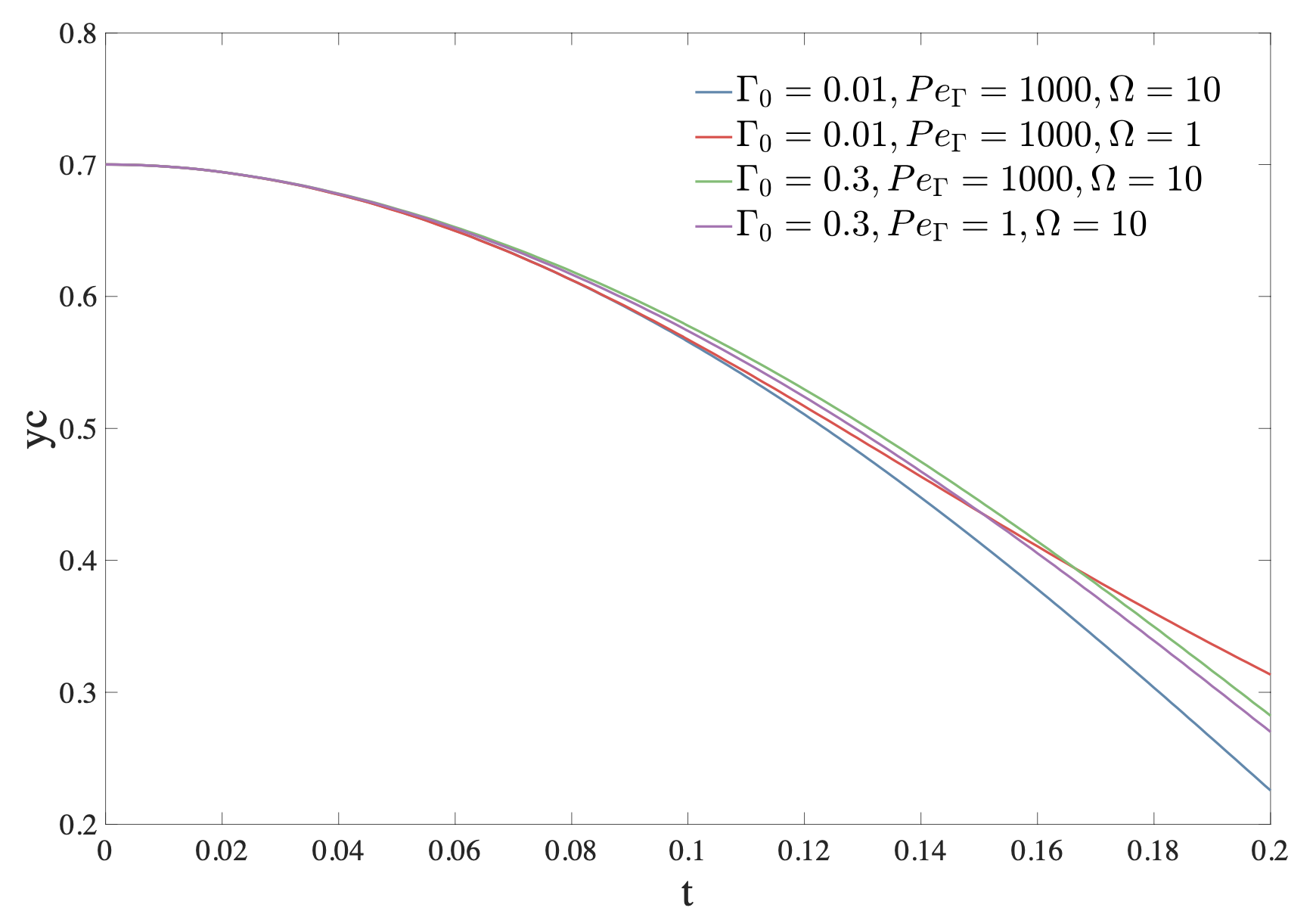} 
\caption{Temporal evolution of the y-coordinate, $y_c$, of the mass center of the drop. The initial particle concentration is $c(t=0) = 0.01$.} \label{figure:yc}
\end{figure}

\begin{figure}[hbt!] 
\centering
\includegraphics[width=0.8\textwidth]{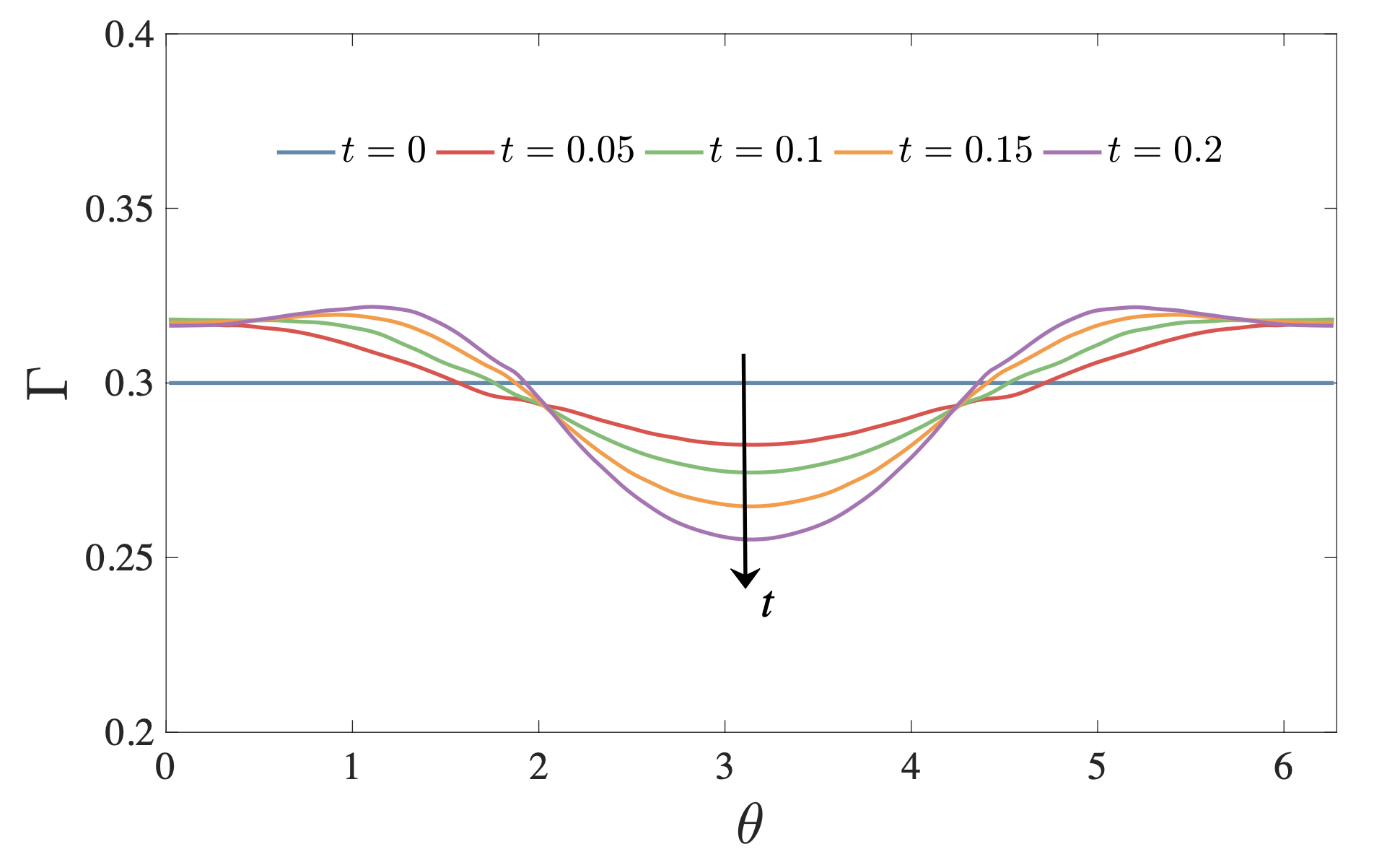} 
\caption{Temporal evolution of the surfactant concentration at the interface. The initial particle concentration is $c(t=0) = 0.01$. $\theta$ is the angle (along clockwise direction) between the positive y direction and the position of the surface relative to the center of the drop ($x_c$, $y_c$). $\theta = 0, 2\pi$ correspond to the top of the drop. $\theta = \pi$ corresponds to the bottom of the drop.} \label{figure:Gamma}
\end{figure}

\begin{figure}[hbt!] 
\centering
\includegraphics[width=0.8\textwidth]{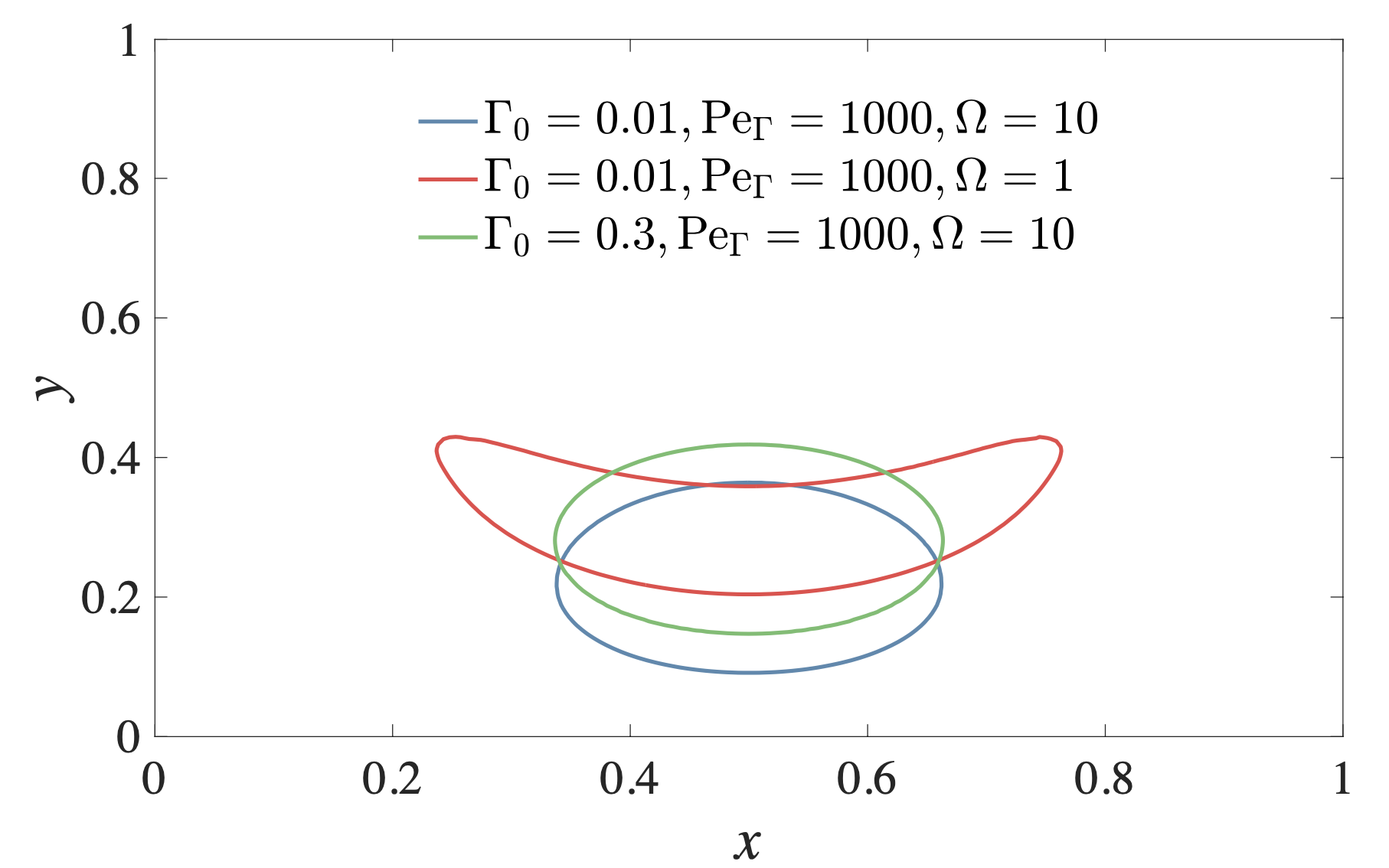} 
\caption{Temporal evolution of the shape of the drop. The initial particle concentration is $c(t=0) = 0.01$.} \label{figure:shape}
\end{figure}

Figure \ref{figure:yc} shows the effects of surfactant properties on $y_c$, which is the y-coordinate of the mass center of the drop. Increasing the Peclet number $\mathrm{Pe}_\Gamma$ (the ratio of the convection of the surfactants to the diffusion) decreases the settling velocity.  This makes sense since the surfactants are advected from the bottom of the drop to the top, as shown in Figure \ref{figure:Gamma}.  For a larger Peclet number $\mathrm{Pe}_\Gamma$, there are more surfactants (smaller surface tension) at top and less surfactants (larger surface tension) at bottom, this surface tension gradient exerts a larger force on the drop which points upward.  Additionally, decreasing the magnitude of the surface tension $\Omega$ (from 10 to 1) is found to decrease the settling velocity, as shown in Figure \ref{figure:yc}. This reduction of settling velocity is caused by the deformation of the drop, as shown in Figure \ref{figure:shape}.

\begin{figure}[hbt!] 
\centering
\includegraphics[width=0.8\textwidth]{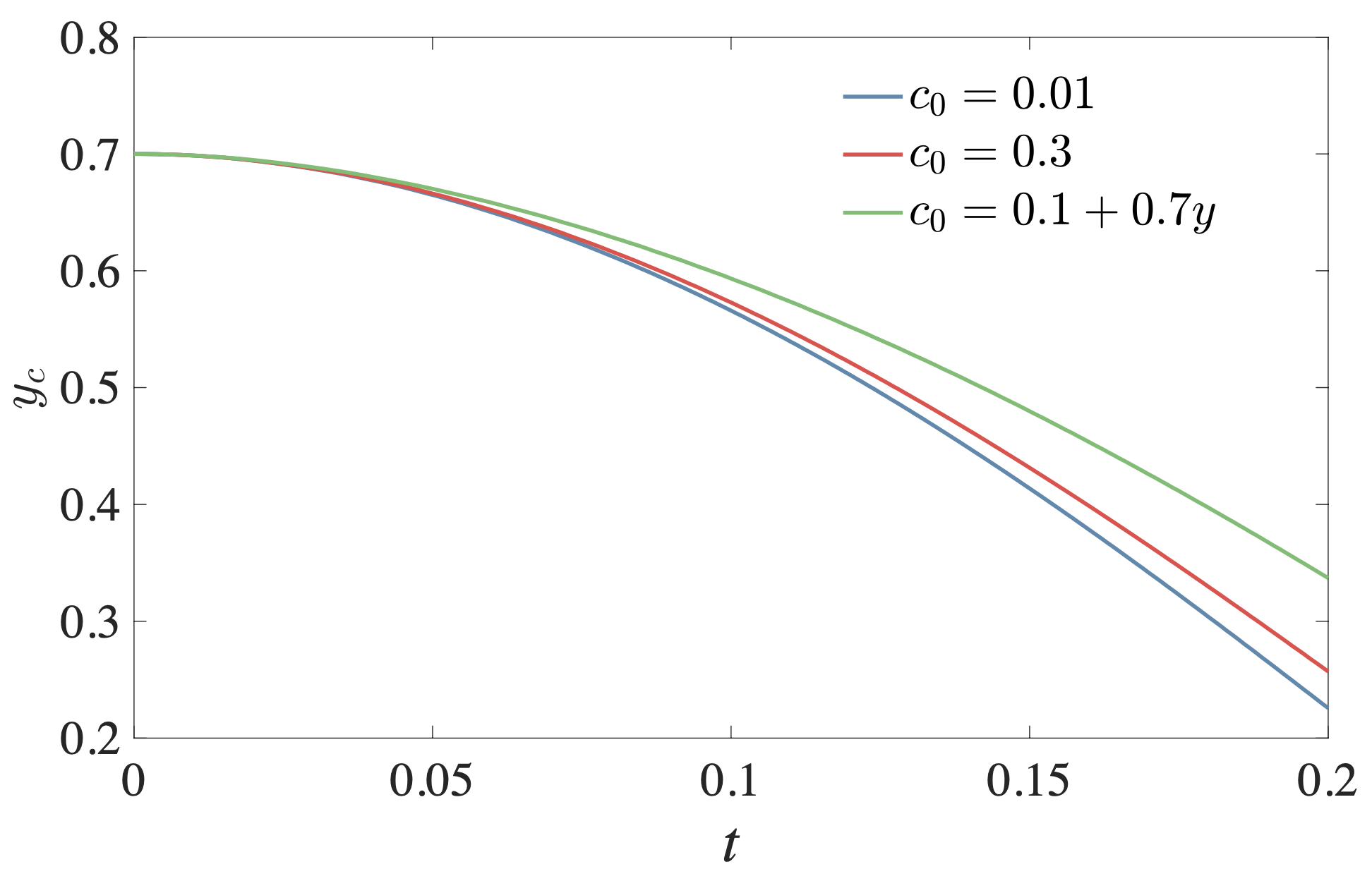} 
\caption{Temporal evolution of the y-coordinate, $y_c$, of the mass center of the drop with respect to different initial particle concentration $c_0$. The initial surfactant concentration is $\Gamma(t=0) = 0.01$.} \label{figure:yc:c}
\end{figure}

\begin{figure}
\centering
{
\begin{minipage}[b]{3in}
\centerline{\includegraphics[width=2.2in]{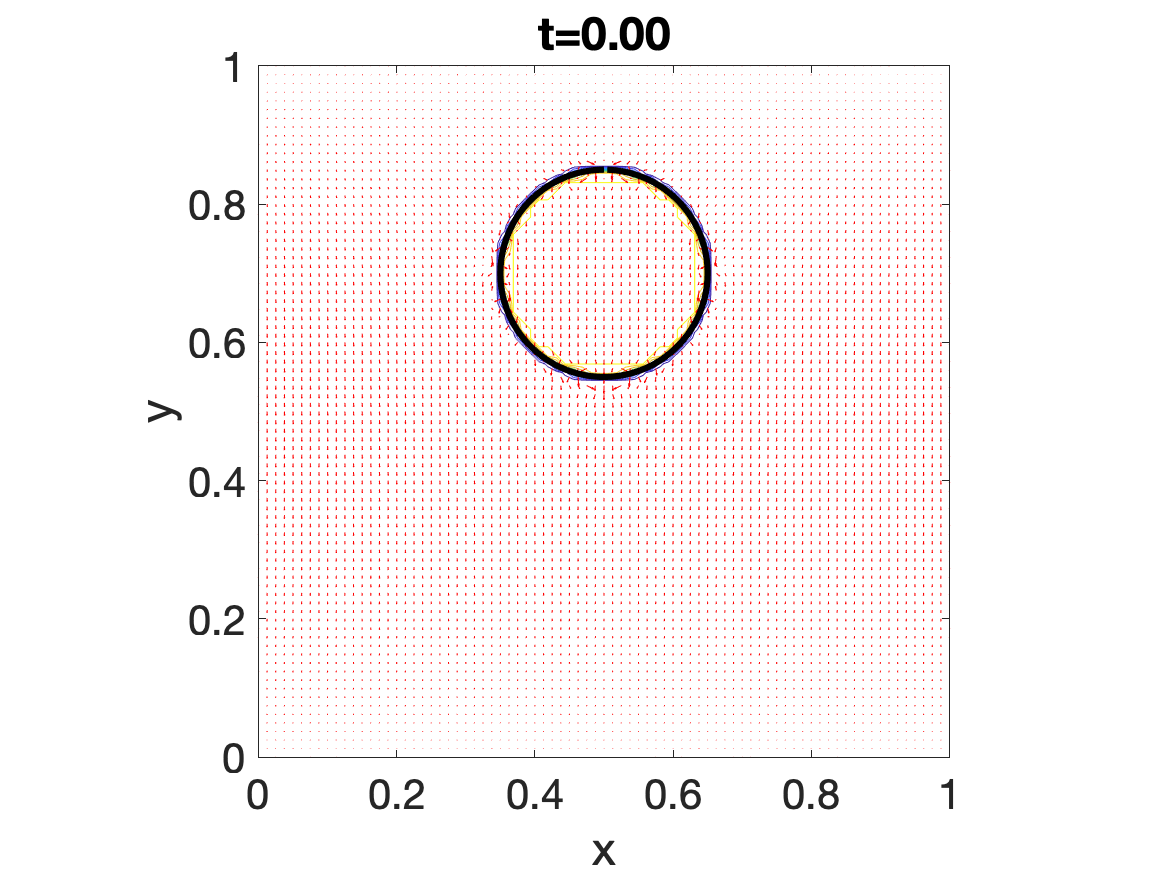}}
\end{minipage}
}
\hfill
{
\begin{minipage}[b]{3in}
\centerline{\includegraphics[width=2.2in]{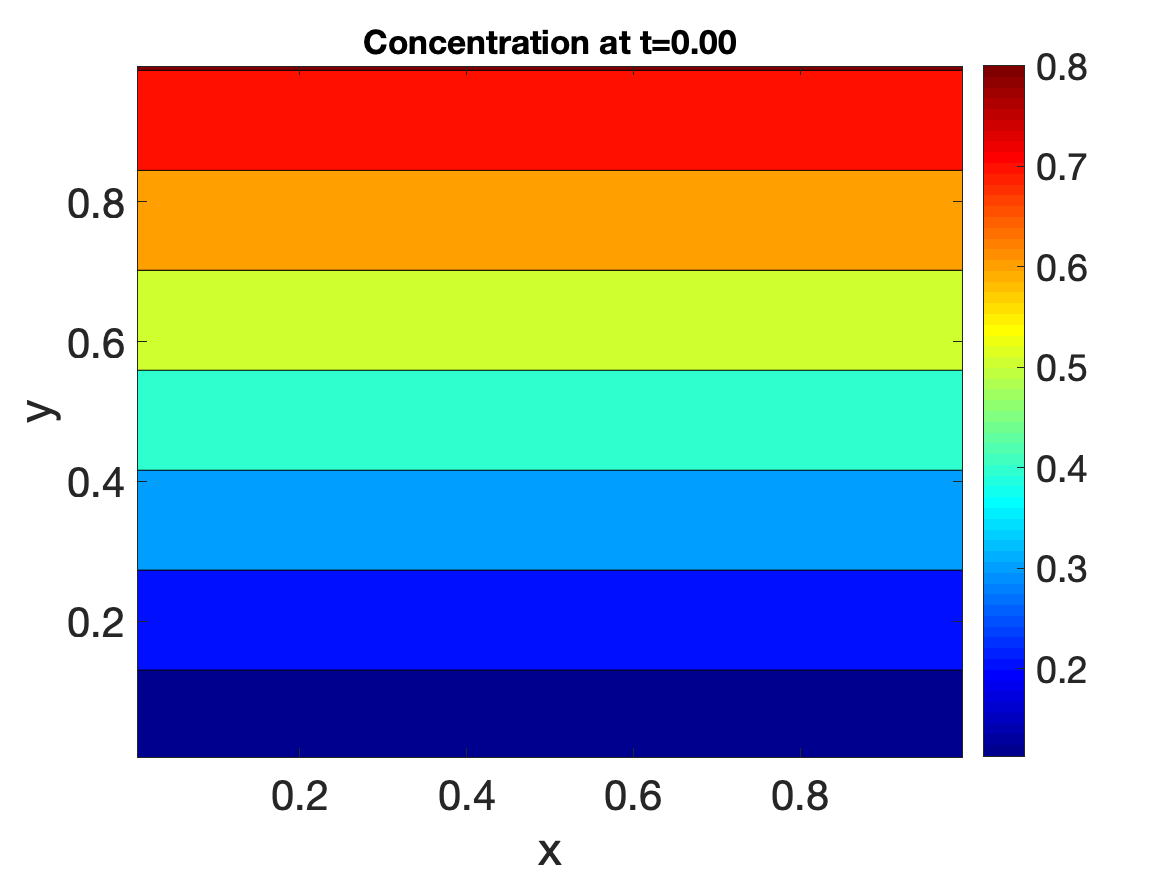}}
\end{minipage}
}
\vfill
{
\begin{minipage}[b]{3in}
\centerline{\includegraphics[width=2.2in]{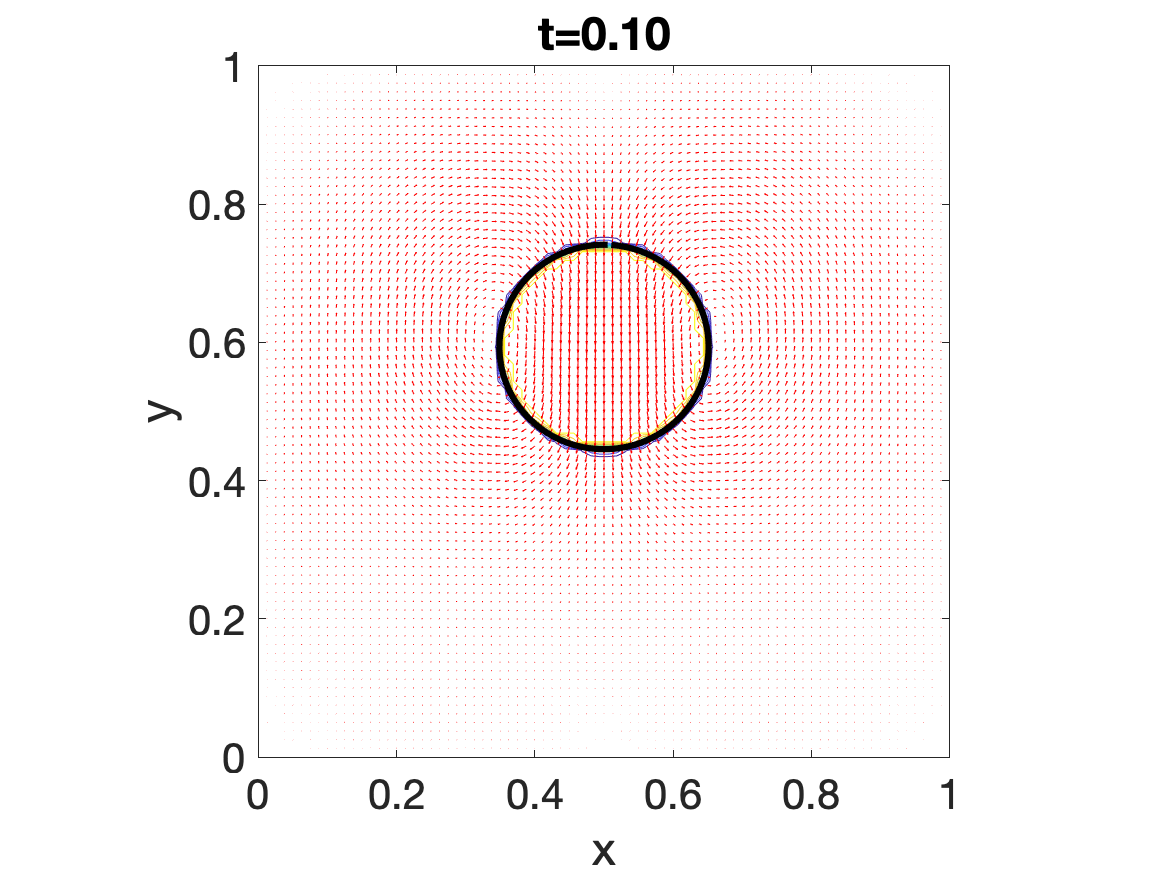}}
\end{minipage}
}
\hfill
{
\begin{minipage}[b]{3in}
\centerline{\includegraphics[width=2.2in]{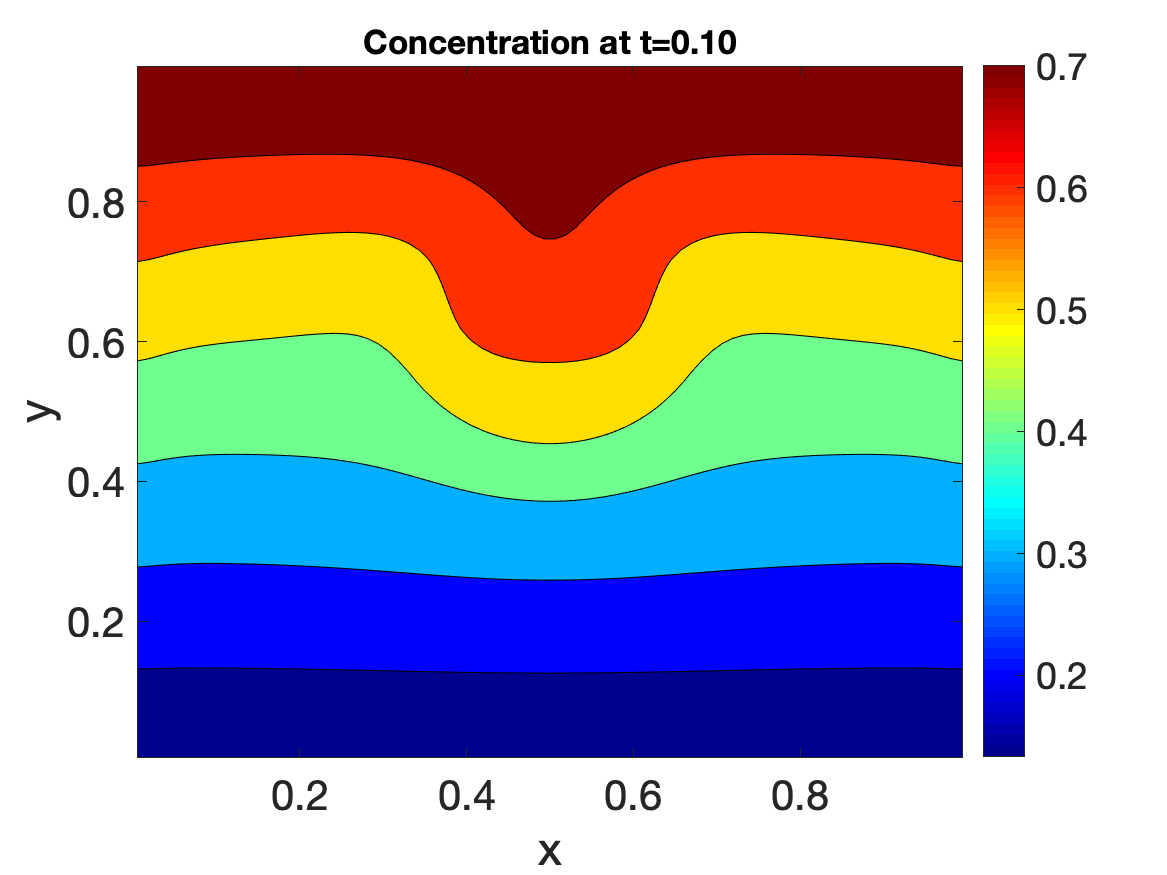}}
\end{minipage}
}
\vfill
{
\begin{minipage}[b]{3in}
\centerline{\includegraphics[width=2.2in]{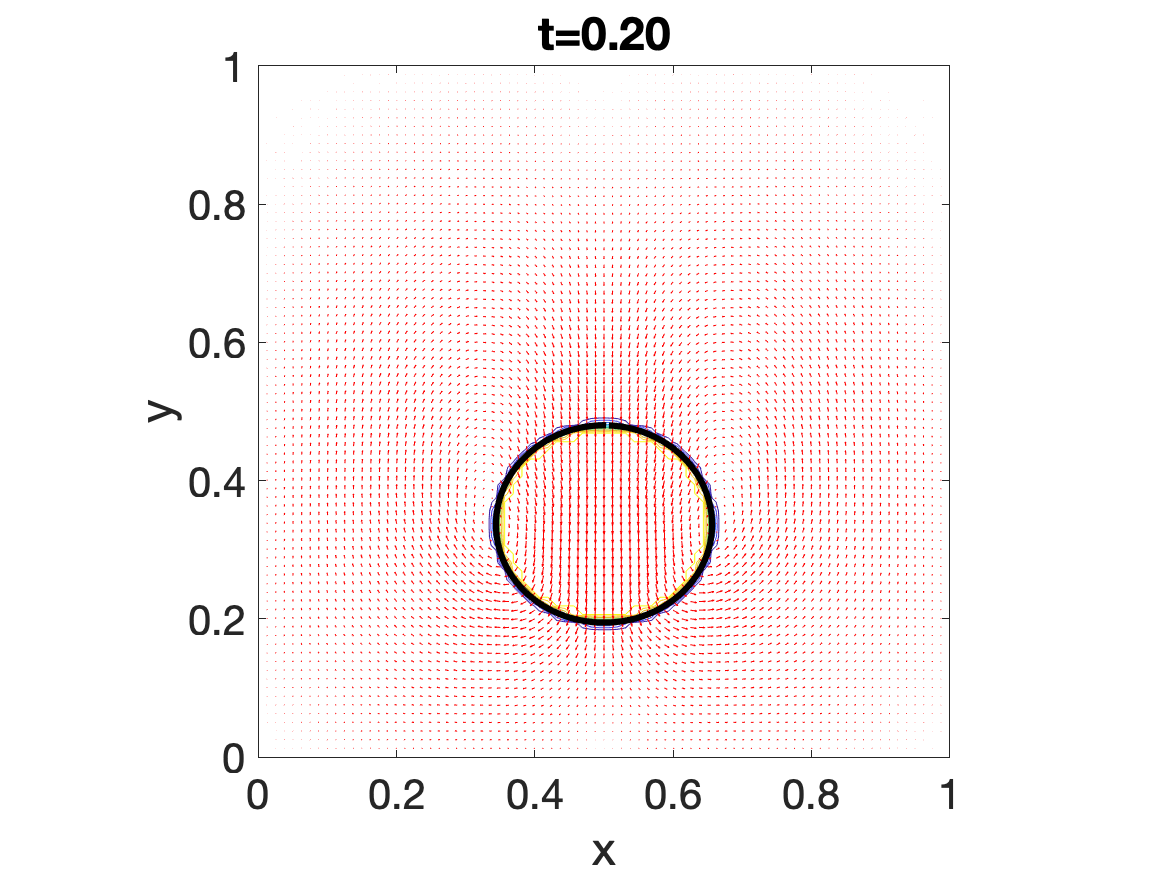}}
\end{minipage}
}
\hfill
{
\begin{minipage}[b]{3in}
\centerline{\includegraphics[width=2.2in]{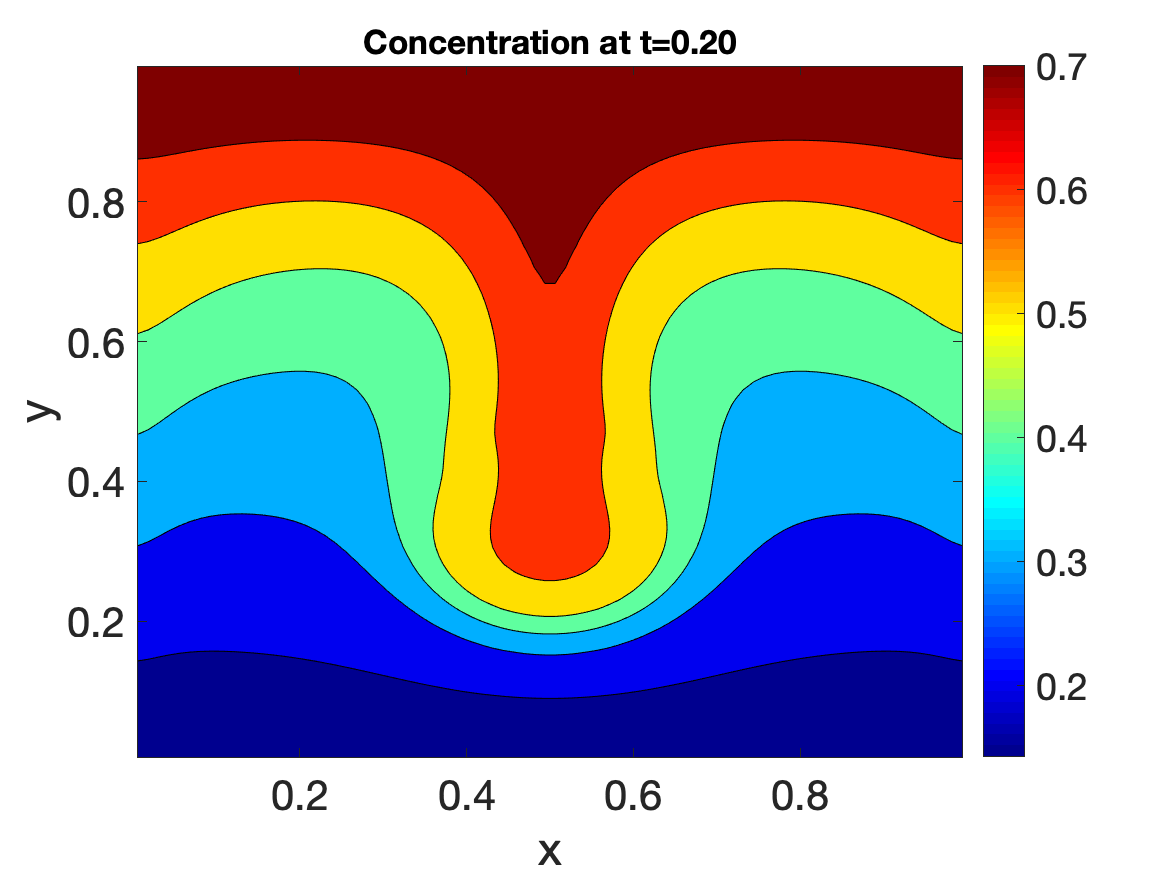}}
\end{minipage}
}
\vfill
{
\begin{minipage}[b]{3in}
\centerline{\includegraphics[width=2.2in]{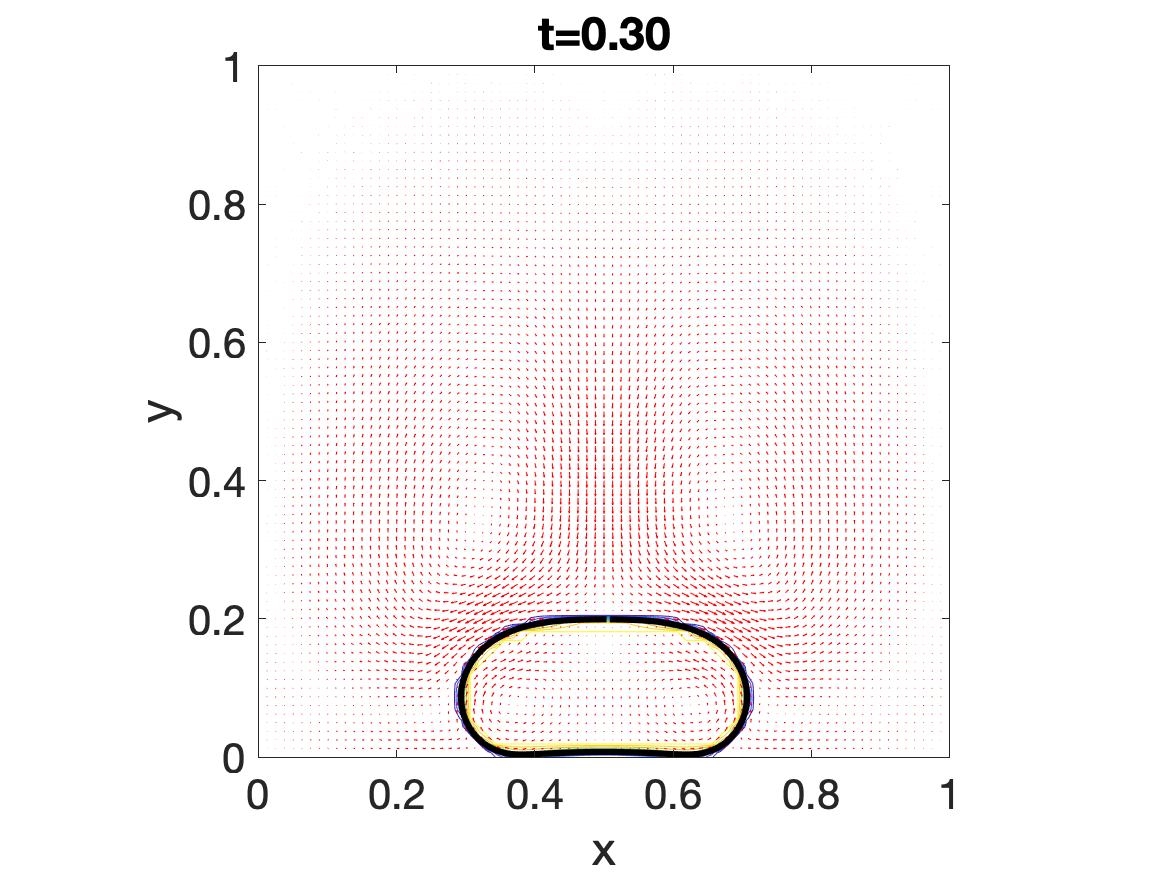}}
\end{minipage}
}
\hfill
{
\begin{minipage}[b]{3in}
\centerline{\includegraphics[width=2.2in]{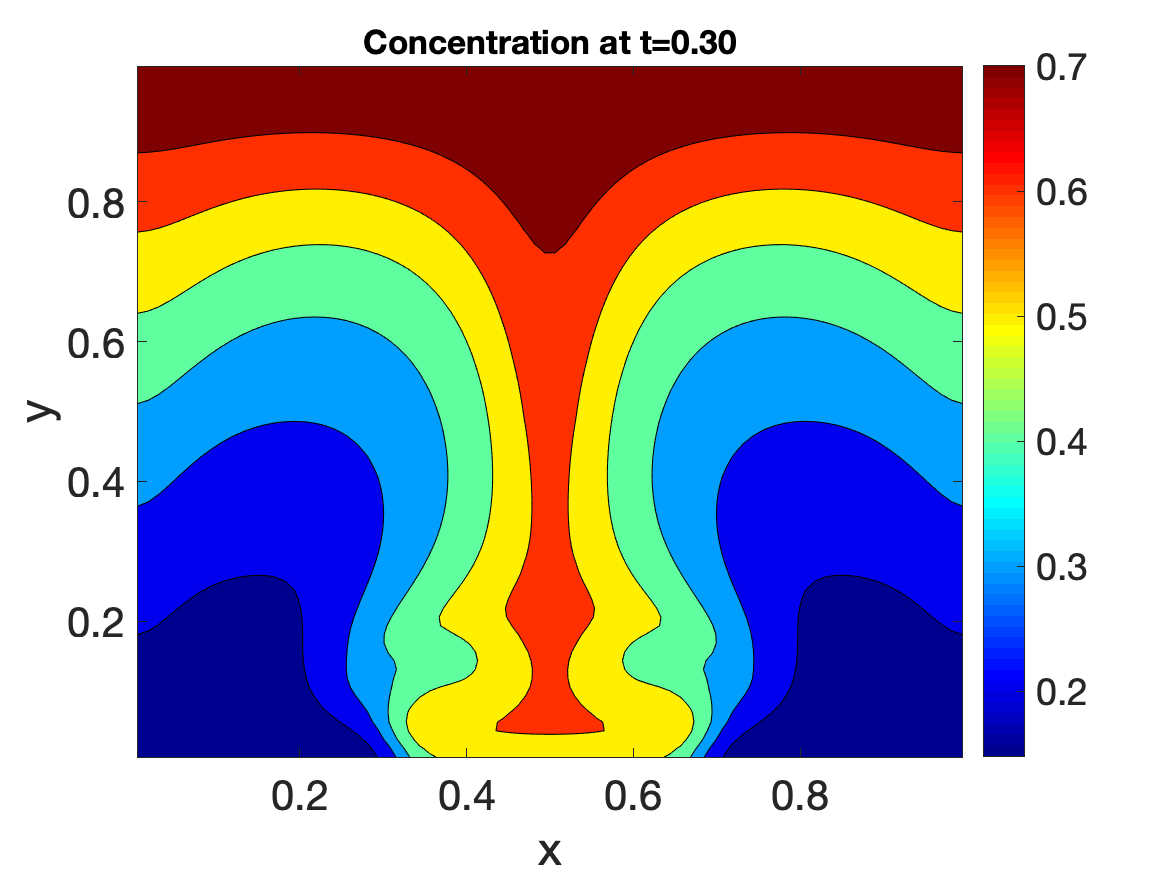}}
\end{minipage}
}
\vfill
{
\begin{minipage}[b]{3in}
\centerline{\includegraphics[width=2.2in]{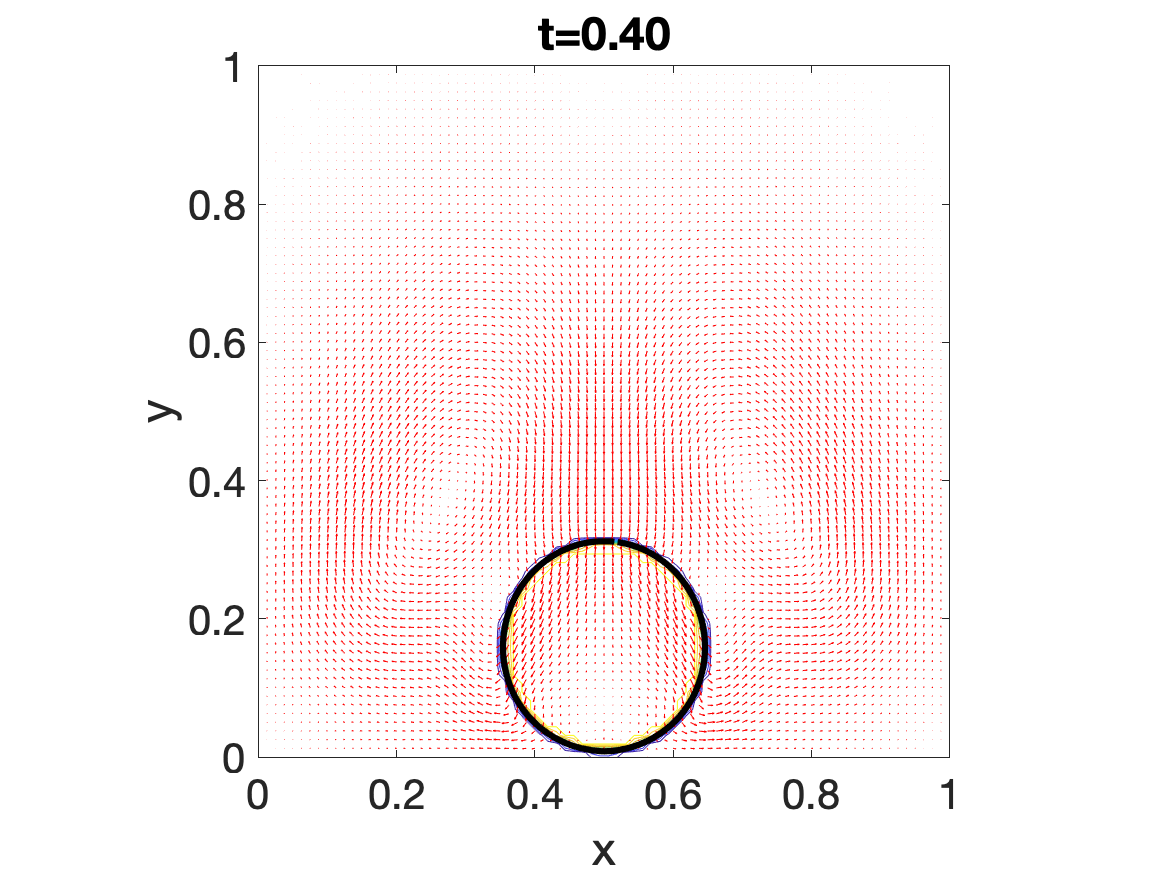}}
\end{minipage}
}
\hfill
{
\begin{minipage}[b]{3in}
\centerline{\includegraphics[width=2.2in]{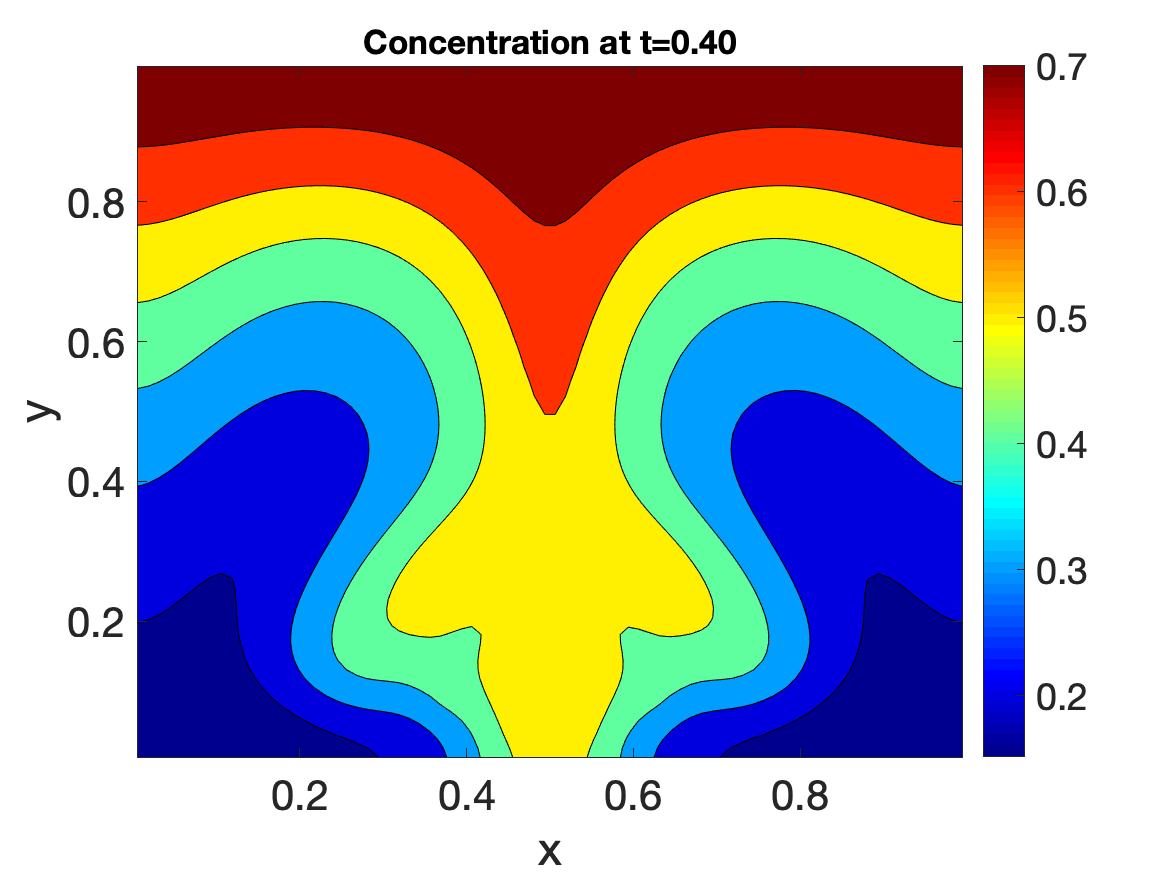}}
\end{minipage}
}

\caption{Temporal evolution of the shape of the drop and contour plots of $c$ with $c(t=0) = c_0 =  0.1+0.7y$, $\Gamma_0 = 0.01$.}
\label{figure:shape:c}
\end{figure}

Figure \ref{figure:yc:c} shows the effects of initial particle concentration on $y_c$.  Increasing local particle concentration increases local viscosity which will delay the settling of the drop. The temporal evolution of the particle concentration and the shape of the drop with $c_0 = 0.1 + 0.7y$ is shown in Figure \ref{figure:shape:c}. Initially, the viscosity in the system increases with $y$. If the drop remains static, the particle concentration $c = c(y)$ will approach 0.4 under diffusion with long enough time. Figure \ref{figure:shape:c} indicates that the settling of drop helps mix the fluid: particles in the domain with high concentration are carried by the drop to the domain with low concentration.

\section{Conclusion}
This project studies the effects of particles and insoluble surfactants on the dynamics of the deformable drop. A 2D simulation is performed with MATLAB. The surfactants not only decrease the surface tension but also induce a surface tension gradient, called Marangoni effect. The surface tension plays a role in maintaining the spherical shape of the drop.  On the settling process, the drop becomes more flat with a smaller surface tension, thus, the setting velocity of the mass center of the drop is smaller, as shown in Figure \ref{figure:shape}.  Additionally, the surfactant migrates from the bottom of the drop to the top under both convection and diffusion.  Increasing the Peclet number of surfactant concentration $\mathrm{Pe}_\Gamma$ enhances convection, i.e., the surface tension gradient is larger, which decreases the setting velocity of the drop, as shown in Figure \ref{figure:shape}. The local viscosity of the drop depends on the local concentration of particles. Increasing particle concentration, i.e., increases the viscosity, decreases the setting velocity of the drop, as shown in Figure \ref{figure:yc:c}.  Also, for a system with initial particle concentration increases with height, the settling of the drop accelerates mixing the system, as shown in Figure \ref{figure:shape:c}.

\section*{Code Availability}
The code can be downloaded by clicking \href{https://github.com/zhongxiaoxu/Computational_methods_interfacial_dynamics.git}{this link.}

\bibliography{sample}

\end{document}